\documentclass[twocolumn,showpacs,amsmath,amssymb,prb,footinbib,floatfix,superscriptaddress]{revtex4-1}

\usepackage{graphicx}% Include figure files
\usepackage{dcolumn}% Align table columns on decimal point
\usepackage{bm}% bold math
\usepackage{helvet}
\usepackage{amssymb}
\usepackage{amsmath}
\usepackage{amsfonts}
\usepackage{hyperref}
\usepackage{color}

\newcommand{\beq}{\begin{equation}}
\newcommand{\eeq}{\end{equation}}

\date{\today}
\begin{document}

\title{Gap scaling at Berezinskii-Kosterlitz-Thouless quantum critical points in one-dimensional Hubbard and Heisenberg models}

\author{M. Dalmonte$^{\S}$}
\email{marcello.dalmonte@uibk.ac.at}
\affiliation{Institute for Quantum Optics and Quantum Information of the Austrian Academy of Sciences, 
and Institute for Theoretical Physics, University of Innsbruck, A-6020 Innsbruck, Austria}
\author{J. Carrasquilla$^{\S}$}
\email{jcarrasquilla@perimeterinstitute.ca}
\address{Perimeter Institute for Theoretical Physics, Waterloo, Ontario, Canada N2L 2Y5}
\author{L. Taddia$^{\S}$}
\email{luca.taddia2@gmail.com}
\thanks{\\$^{\S}$ These authors contributed equally to this work.}
\address{Scuola Normale Superiore, Piazza dei Cavalieri 7, 56126 Pisa, Italy, \\ 
and CNR-INO, UOS di Firenze LENS, Via Carrara 1, 50019 Sesto Fiorentino, Italy}
\author{E. Ercolessi}
\affiliation{Dipartimento di Fisica e Astronomia, Universit\`a di Bologna and INFN, 
via Irnerio 46, 40127 Bologna, Italy}
\author{M. Rigol}
\address{Department of Physics, The Pennsylvania State University, University Park, PA 16802, USA}

%%%%%%
\begin{abstract}
We discuss how to locate critical points in the Berezinskii-Kosterlitz-Thouless (BKT) universality 
class by means of gap-scaling analyses. While accurately determining such points using gap extrapolation 
procedures is usually challenging and inaccurate due to the exponentially small value of the gap 
in the vicinity of the critical point, we show that a generic gap-scaling analysis, including the 
effects of logarithmic corrections, provides very accurate estimates of BKT transition points in 
a variety of spin and fermionic models. As a first example, we show how the scaling procedure, 
combined with density-matrix-renormalization-group simulations, performs extremely well in a 
non-integrable spin-$3/2$ XXZ model, which is known to exhibit strong finite-size effects. 
We then analyze the extended Hubbard model, whose BKT transition has been debated, 
finding results that are consistent with previous studies based on the scaling of the Luttinger-liquid
parameter. Finally, we investigate an anisotropic extended Hubbard model, for which we present the first 
estimates of the BKT transition line based on large-scale density-matrix-renormalization-group
simulations. Our work demonstrates how gap-scaling analyses can help to locate accurately and 
efficiently BKT critical points, without relying on model-dependent scaling assumptions. 
\end{abstract}

%05.70.Jk: Critical Phenomena
%71.10.Pm: Fermions in reduced dimensions
%75.10.Pq: Spin chain models
%71.10.Fd: Lattice fermion models (Hubbard model, etc.)
%\pacs{05.30.Rt, 64.70.Tg, 03.67.Mn, 05.70.Jk}

%%%%%%
\maketitle
%%%%%

\section{Introduction}

Quantum phase transitions in one-dimensional (1D) systems are one of the most remarkable consequences 
of the enhanced role of quantum fluctuations in reduced dimensions. \cite{Giamarchi2003,bosonization,RevModPhys.83.1405} 
While systems with discrete symmetries, such as the Ising model, can still undergo phase transitions associated with the 
spontaneous breaking of those symmetries, 1D systems with continuous symmetries, such as spin-rotation 
or particle-number conservation, cannot get spontaneously broken under rather general conditions 
due to the Mermin-Wagner-Hohenberg theorem.\cite{Mermin1966,Hohenberg1967} Similarly to classical two-dimensional 
systems at finite temperature, 1D quantum systems endowed with continuous symmetries 
can still undergo a quantum phase transition according to the Berezinskii-Kosterlitz-Thouless (BKT) 
mechanism. \cite{Berezinskii1972,Kosterlitz1973} Such transitions play a key role in the physics of 1D spin, bosonic, and 
fermionic models, which find incarnations as diverse as different magnetic 
compounds,~\cite{RevModPhys.83.1405,giamarchi2012} and ultracold 
atom and molecule gases trapped in optical lattices.\cite{RevModPhys.83.1405,RevModPhys.80.885,Hadzibabic2006,Haller2010} 

In the BKT scenario, the phase transition point is conformal, and in its vicinity the gap closes exponentially 
as a function of the microscopic parameters.\cite{Itzykson1989} This feature makes numerical 
investigations of BKT transitions challenging, as very large systems sizes are required 
in order to avoid severe finite-size effects. Usually, techniques from field theory can be used to 
pin down the transition point. These include methods that use correlation functions to 
track the scaling dimension of relevant operators close to transition 
point,~\cite{Giamarchi2003,PhysRevLett.92.236401,Rachel:2012jk} or entanglement 
entropies to monitor the behavior of the central charge of the 
system.\cite{1742-5468-2004-06-P06002,Laeuchli:2008qy} While these approaches have 
provided notable insights in the context of several lattice models, it is highly desirable to 
develop and benchmark alternative methods based on the spectral properties, which do not rely 
on any {\it a priori} knowledge of the underlying field theory, and at the same time can cope 
well with logarithmic corrections. Moreover, gap-based methods are, from the computational side, 
potentially less demanding than evaluating correlations functions, and precise bounds on the error 
can be given when employing variational techniques such as the density-matrix-renormalization-group 
(DMRG).\cite{DMRG1,DMRG2,schollwock_review_05}

Here, we show how refined gap-scaling analyses provide accurate insights on phase diagrams of 
1D spin and fermionic models undergoing a BKT transition. Our technique relies on a recently 
proposed scaling ansatz for the gap close to the critical point, which was successfully applied 
to the $t$-$V$-$V'$ model of spinless fermions in 1D (equivalent to the spin-1/2 XXZ chain 
with next-nearest neighbor $S^zS^z$ interactions)\cite{Mishra2011} and the 1D Bose-Hubbard 
model.\cite{Carrasquilla2013} All our calculations are done using DMRG, which allows us 
to accurately and efficiently determine the gaps needed for the scaling analyses.

As a first step in our study, in Sec.~\ref{Sec:spin}, we apply the scaling approach to an $S=3/2$ 
XXZ chain, where the exact location of the BKT point is known, but it is difficult 
to pinpoint numerically owing to strong finite-size corrections. Using simulations with both 
periodic and open boundary conditions, we show that the scaling method is able to locate the transition 
point with errors at the $\sim$1\% level in the presence of strong logarithmic corrections (for periodic boundary 
conditions). In Sec.~\ref{Sec:EHM}, we discuss the feasibility of our approach for extended 
Hubbard models including nearest-neighbor interactions, where the existence and location of a BKT transition 
separating a spin-density-wave and a bond-ordered phase has been extensively 
debated.\cite{PhysRevLett.92.246404,PhysRevB.83.075111,PhysRevLett.99.216403,Tsuchiizu2004,Tam2006,
Nakamura2000,Jeckelmann2002,Barbiero:2013lq,Glocke:2007rz,Sengupta2002} In 
Sec.~\ref{Sec:AEHM}, we investigate an anisotropic version of the EHM, the anisotropic-extended-Hubbard 
model, where spin-rotation symmetry is explicitly broken. For the latter model, we complement 
the gap scaling analysis with a correlation-function method based on the scaling of the Luttinger 
parameter, which provides an independent way to locate the transition point. Finally, we 
recapitulate the main results and discuss possible extensions of our work in Sec.~\ref{Sec:concl}. 

\section{Spin-$3/2$ XXZ chain}\label{Sec:spin}

Spin chains are prototypical models of one-dimensional (1D) quantum systems.\cite{RevModPhys.83.1405} 
The first spin chains introduced were of the Heisenberg (also known as XXX) type:\cite{Heisenberg1928}
\begin{equation}\label{XXX}
\hat{H}=J\sum_{j}\vec{S}_j\cdot\vec{S}_{j+1},
\end{equation}
where $J\in\mathbb{R}$ and $\vec{S}$ are matrices belonging to some finite-dimensional 
representation of SU(2). In the antiferromagnetic case, $J>0$, and for general finite-dimensional 
representations of SU(2), Haldane~\cite{Haldane1,Haldane2,Haldane3} conjectured that the ground 
state should be gapped for integer values of $S$ and gapless (belonging to the SU(2)$_1$ 
Wess-Zumino-Novikov-Witten universality class) for half-integer values of $S$. This conjecture 
has been extensively verified analytically and numerically 
(see, e.g., Refs.~\onlinecite{ConjVerAna1,ConjVerAna2,Hallberg1996}). 

The XXZ chain, on the other hand, is a generalization of the Heisenberg one that is 
obtained by introducing anisotropy along one, namely the $z$-, axis. The Hamiltonian can be cast 
in the form
\begin{equation}
\hat{H}=-J\sum_{j}\left(S_j^xS_{j+1}^x+S_j^yS_{j+1}^y-J_z S_j^zS_{j+1}^z\right),
\end{equation}
where $J_z$ is the anisotropy coefficient (we set $J=1$ as our energy unit in what follows). 
For half-integer $S$, $J_z=1$ is a critical point separating a conformal phase 
(a Tomonaga-Luttinger liquid, $-1<J_z<1$) from a N\'eel phase ($J_z>1$)~\cite{schulz1986,affleck1987,Hallberg1996}. 
This phase transition is known to belong to the BKT universality class, and, in the vicinity of the 
critical point, the low-energy spectrum is well-described by a sine-Gordon 
model:\cite{haldane1981,bosonization,Giamarchi2003}
\begin{equation}\label{H_TLL}
 \hat{H}=\int dx\left\{ \frac{v_s}{2\pi}\left[\frac{(\partial_x \vartheta)^2}{K}
 +K(\partial_x\varphi)^2\right] + g\cos[\sqrt{4\pi}\varphi]\right\}\nonumber,
\end{equation}
where $v_s$ is the sound velocity, $\vartheta$ and $\varphi$ are conjugated density and phase bosonic 
fields, $K$ is the Luttinger-liquid parameter, related to the compactification radius
$R_\vartheta= 1/R_\varphi$ of the fields via $K=1/(4\pi R_\vartheta^2)$,\cite{bosonization} and 
the last term gives rise to a finite mass in the spectrum for $K<1/2$.

While for the $S=1/2$ integrable case numerical methods work relatively well locating $J_z=1$ as the 
transition point,\cite{bosonization,PhysRevB.84.085110} strong logarithmic corrections arise for 
$S\geq 3/2$, making the numerical detection of the BKT point difficult.\cite{Dalmonte2012} In certain 
cases precise knowledge of the underlying logarithmic scaling of $K$ in terms of perturbed conformal field 
theories can be provided, making methods based on correlation-function applicable.\cite{Affleck1989,Hallberg1996} 
As the latter rely on {\it ad hoc} assumptions based on the symmetry content of the theory at the critical point, 
they cannot, in general, be extended to other models. 

\begin{figure}[b]
\includegraphics[width=0.75\columnwidth]{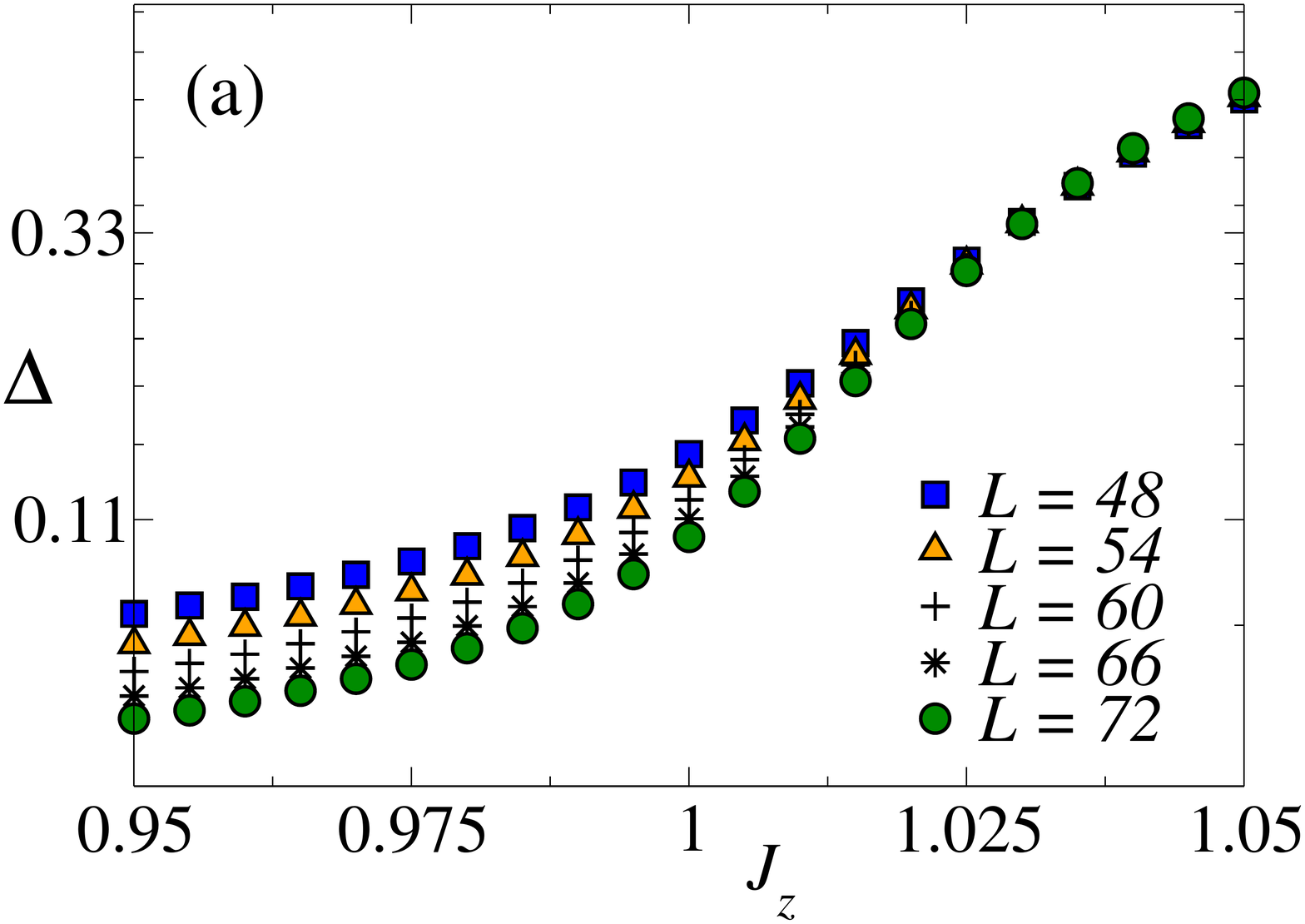}
\includegraphics[width=0.75\columnwidth]{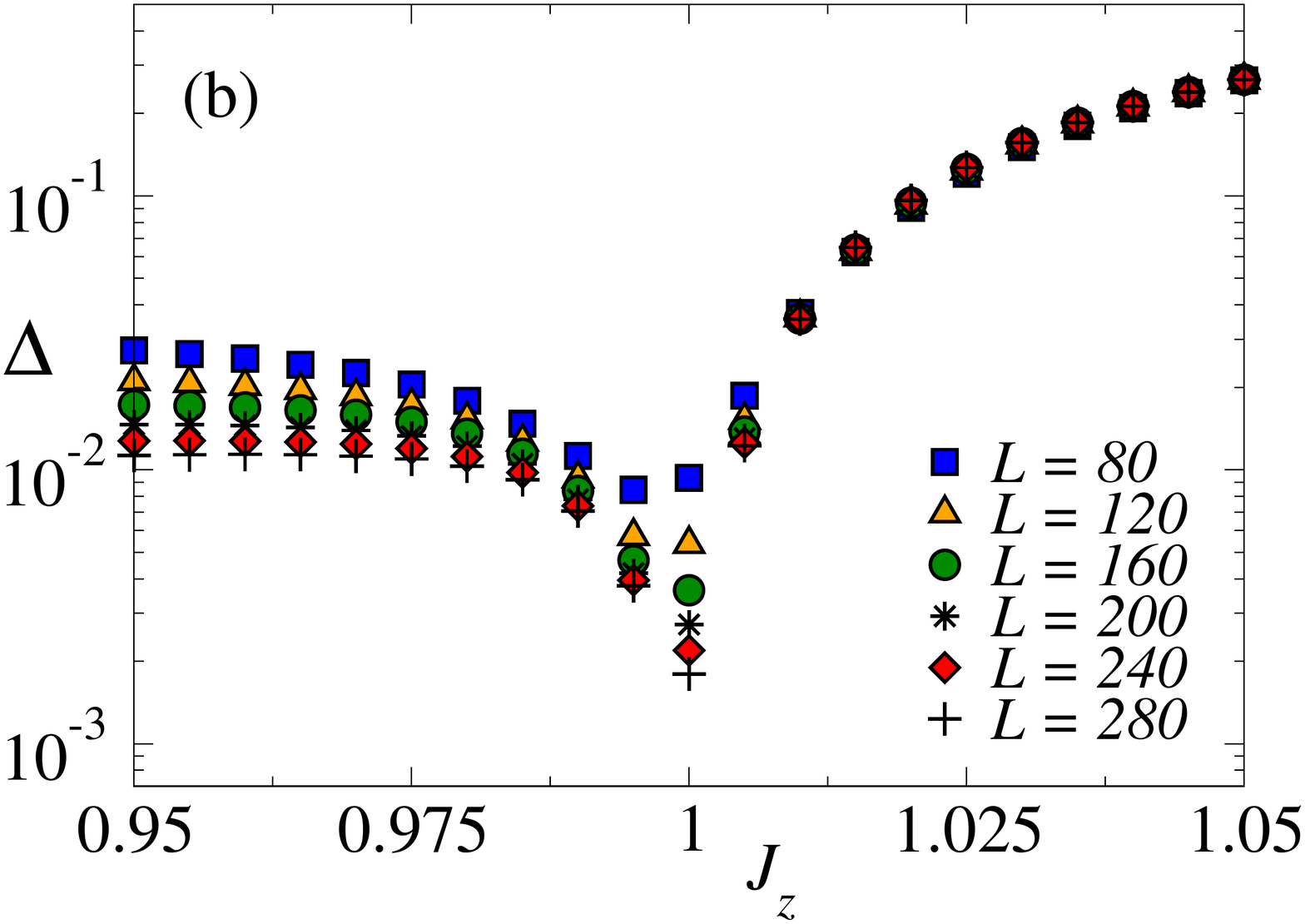}
 \caption{(Color online) Energy gaps for the spin-$3/2$ XXZ chain with PBCs (a) and OBCs (b), 
as a function of $J_z$ and for different values of $L$. In both cases, the gap axis has a 
logarithmic scale for sake of clarity.
}\label{XXZ_gap}
\end{figure}

It is our aim in the following to 
investigate how gap scaling methods, which do not rely on field theoretical assumptions, can be employed 
to locate BKT transitions points. In this context, the $S=3/2$ XXZ model represents an ideal 
benchmark, since, on the one hand, the location of the transition point is known, and, on the other 
hand, strong logarithmic corrections need to be incorporated, providing a strict test for the 
reliability of the method itself. This scaling analysis used here was applied to the $S=1/2$ integrable 
case in Ref.~\onlinecite{Mishra2011}, and the critical point was found to be at $J^c_z=1.01\pm0.005$, 
in agreement with the exact result.

\begin{figure*}[t]
\includegraphics[width=0.65\linewidth]{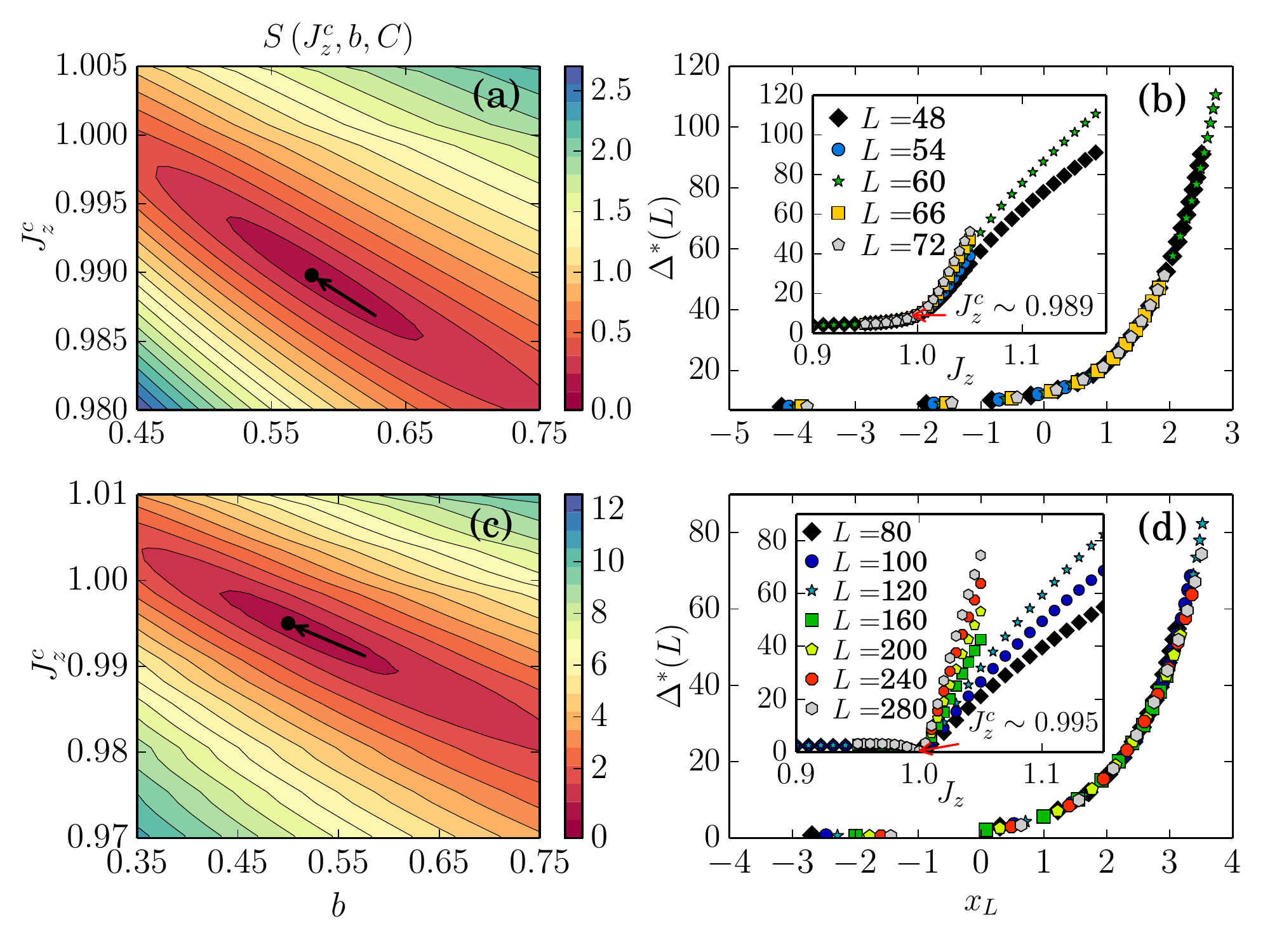}
\caption{(Color online) (a) Contour plot of the sum of squared residuals $S(J^c_{z},b,C)$ for the XXZ
model with PBCs. The arrow signals the location of the minimum value of $S$. The black lines are equally
spaced contour lines where $S$ is constant. (b) Best collapse of the data for $\Delta^*(L)$ vs $x_L$
corresponding to $J^c_{z}=0.989$, $b = 0.58$, and $C=-4.5$. The inset shows the rescaled gap vs $J_z$.
A similar analysis for the system with OBCs is presented in panels (c) and (d). $J_z$ and $\Delta^*(L)$
are presented in units of $J$, while $b$ and $S$ are shown in units of $J^{1/2}$ and $J^2$,
respectively.}\label{32XXY_sc}
\end{figure*}

Following Refs.~\onlinecite{Mishra2011,Carrasquilla2013}, we locate the critical point of the BKT 
transition in the $S=3/2$ case by studying the scaling of the spin excitation gap 
$\Delta\left(L\right) = E_0(L) - E_1(L)$ on finite systems, where $E_{p}(L)$ is the ground state 
energy at size $L$ in the magnetization sector $\sum_{j}S^z_j = p$. We have performed DMRG simulations 
both with periodic (PBCs) and open (OBCs) 
boundary conditions: in the former (latter) case, we consider system sizes in the range 
$L\in[48,72]$ ($[60,280]$), keeping up to $m=1200$ (768) Schmidt states in the finite-size sweeping 
procedure. We have performed 6 (4) finite-size sweeps and achieved a truncation error less, in all cases, 
than $10^{-5}$ ($10^{-8}$). The typical errors in the gap are of order $2\times10^{-4} (10^{-6})$ for PBCs (OBCs).
By comparing the results for different boundary conditions, we gain 
insights on the effects of translational invariance, which is broken under OBCs. In both cases, we 
consider anisotropies in the range $J_z\in[0.95,1.05]$. In Figs.~\ref{XXZ_gap}(a) and \ref{XXZ_gap}(b), 
we plot the energy gap $\Delta\left(L\right)$ as a function of $J_z$, for different system sizes $L$, 
and for PBCs and OBCs, respectively. In the chains with PBCs [Fig.~\ref{XXZ_gap}(a)], the gap exhibits 
a rather smooth behavior with changing $J_z$ and increasing $L$. For OBCs [Fig.~\ref{XXZ_gap}(b)], a 
dip occurs very close to the critical point. As expected, under both boundary conditions, the  
magnitude of the gap decreases as $L$ increases for $J_z<1$.

The method described in Ref.~\onlinecite{Mishra2011,Carrasquilla2013} is based on the following 
ansatz for the scaling of the gap in the vicinity of the phase transition,
\begin{equation}
L \Delta\left(L\right) \times \left(1+\frac{1}{2\ln{L}+C} \right)= F\left( \frac{\xi}{L} \right),
\label{eq:scaling}
\end{equation}
where $F$ is a scaling function, $C$ is a nonuniversal constant to be determined, and $\xi$ is the 
correlation length. 
This scaling ansatz is an attempt to include known logarithmic corrections to 
the gap in a BKT transition\cite{Eckle1997} and reduces to the analog relation for the
resistance (which also vanishes exponentially) in the charge-unbinding transition of the two-dimensional 
classical Coulomb gas at the critical point.\cite{PhysRevB.51.6163} In the latter case, the universal 
nature of the coefficient in front of $\ln(L)$ can be traced back to the logarithmic corrections of 
the Weber-Minnhagen finite-size scaling relation of the dielectric function, which is, in turn, 
related to the superfluid stiffness (see Refs.~\onlinecite{PhysRevB.51.6163} and \onlinecite{PhysRevB.37.5986}). 
For isotropic chains, conformal field theory calculations show that the prefactor in front 
of the logarithm is universal and equals 2.~\cite{Affleck1989,Nomura:nr,0305-4470-19-17-008}
Note that in a BKT transition, the correlation length diverges as $\xi\sim\Delta^{-1}\sim\exp(b/\sqrt{J_z-J^c_{z}})$, 
where $b$ is independent of $J_z$. Because of the divergence of the correlation length at the critical point, 
the function $F(\xi/L)$ becomes system-size independent, and thus the data for the rescaled gap 
$\Delta^{*}\left(L\right)\equiv L \Delta\left(L\right) \left[1+1/\left(2\ln{L}+C\right) \right]$ 
for different system sizes $L$ will be independent of $L$ for $J_z\leq J^c_{z}$. Additionally, plots 
of $\Delta^{*}\left(L\right)$ vs $\xi/L$ should collapse onto a unique curve representing $F$. In 
order to plot the scaling collapse, one can rewrite the relation in Eq.~\eqref{eq:scaling} by 
taking the logarithm of the argument of $F$ and considering an alternative function $f$ with 
argument $x_L=\ln L -\ln \xi$. 

We determine the critical point by adjusting the parameters $J^c_{z}$, $b$, and $C$ such that the 
best collapse of the curves $\Delta^{*}\left(L\right)$ vs $x_L$ is obtained. To do that, we represent 
$f$ through an arbitrary high-degree polynomial and fit it on a dense grid of values of $J^c_{z}$, $b$, 
and $C$, to the calculated values of $\Delta^{*}\left(L\right)$ and $x_L$. The desired parameters 
$J^c_{z}$, $b$, and $C$ are selected by minimizing the sum of squared residuals $S(J^c_{z},b,C)$ of 
the fit. We ensure that the results are robust to the choice of polynomial and the interval of 
values of $J_z$ used in the fits. 

The results of this procedure applied to the data in Fig.~\ref{XXZ_gap} are summarized in Fig.~\ref{32XXY_sc}. 
In Fig.~\ref{32XXY_sc}(a), we present a density plot corresponding to $S(J^c_{z},b,C)$ for the XXZ model 
with PBCs, which exhibits a clear minimum at $J^c_{z}=0.989\pm0.01$, $b=0.58\pm0.04$, and $C=-4.5\pm0.2$. In 
Fig.~\ref{32XXY_sc}(b), we display $\Delta^{*}\left(L\right)$ vs $x_L$ for the given set of parameters 
that minimize $S(J^c_{z},b,C)$. The data clearly collapses onto a unique curve representing the function 
$f$. The sensitivity of the results to the selection of the interval of values of $J_z$ used in the fit 
is also included in the error bars such that our results are independent of its choice. In the inset, the 
curves for the rescaled gap vs $J_z$ and different system sizes merge around the critical value $J^c_{z}$ 
found through the minimization procedure. This indicates that the ansatz in Eq.~\eqref{eq:scaling} describes 
well the numerical data around the critical point.

In Fig.~\ref{32XXY_sc}(c), we present $S(J^c_{z},b,C)$ for the XXZ model with OBCs, which exhibits a minimum 
at $J^c_{z}=0.995\pm0.004$, $b=0.50\pm0.02$. In this case, the values of $C$ that minimize $S$ are arbitrarily 
large; in practice this means that logarithmic corrections to the gap are suppressed when OBCs are used. 
In Fig.~\ref{32XXY_sc}(d), we show $\Delta^{*}\left(L\right)$ vs $x_L$ for the given set of parameters that 
minimize $S(J^c_{z},b,C)$. As in the case with PBCs, the data is seen to collapse to a unique curve. In 
this case, it is also verified that the curves of the rescaled gap vs $J_z$ merge around the critical point 
retrieved from the minimization procedure. 

Our results for the critical anisotropy coefficient $J^c_{z}$, both for OBCs and PBCs, are very close to the 
analytical result, indicating that the data and critical behavior of the gap are well described by 
Eq.~\eqref{eq:scaling}. The use of OBCs has clear advantages. First, from the DMRG perspective, the use of 
OBCs generally allows simulating larger system sizes while keeping lower errors in the energy gaps: for the 
same accuracy obtained keeping $m$ states and OBCs, one requires of the order of $m^2$ states in the case of 
PBCs. Second, for the present model, the logarithmic corrections are suppressed when OBCs are used, thus 
effectively reducing the number of parameters that need to be determined in the minimization of $S$. 
While the precise reason for this suppression is not known, we note that the same behavior occurs in
the Bose-Hubbard model~\cite{Carrasquilla2013}. We speculate that, for those models, certain 
marginal operator contribution vanishes at the transition point, akin to what happens at the BKT point 
of the Majumdar-Ghosh chain.~\cite{eggert1996}

\section{One-dimensional extended Hubbard model}\label{Sec:EHM}

\begin{figure}[!b]
\includegraphics[width=0.75\columnwidth]{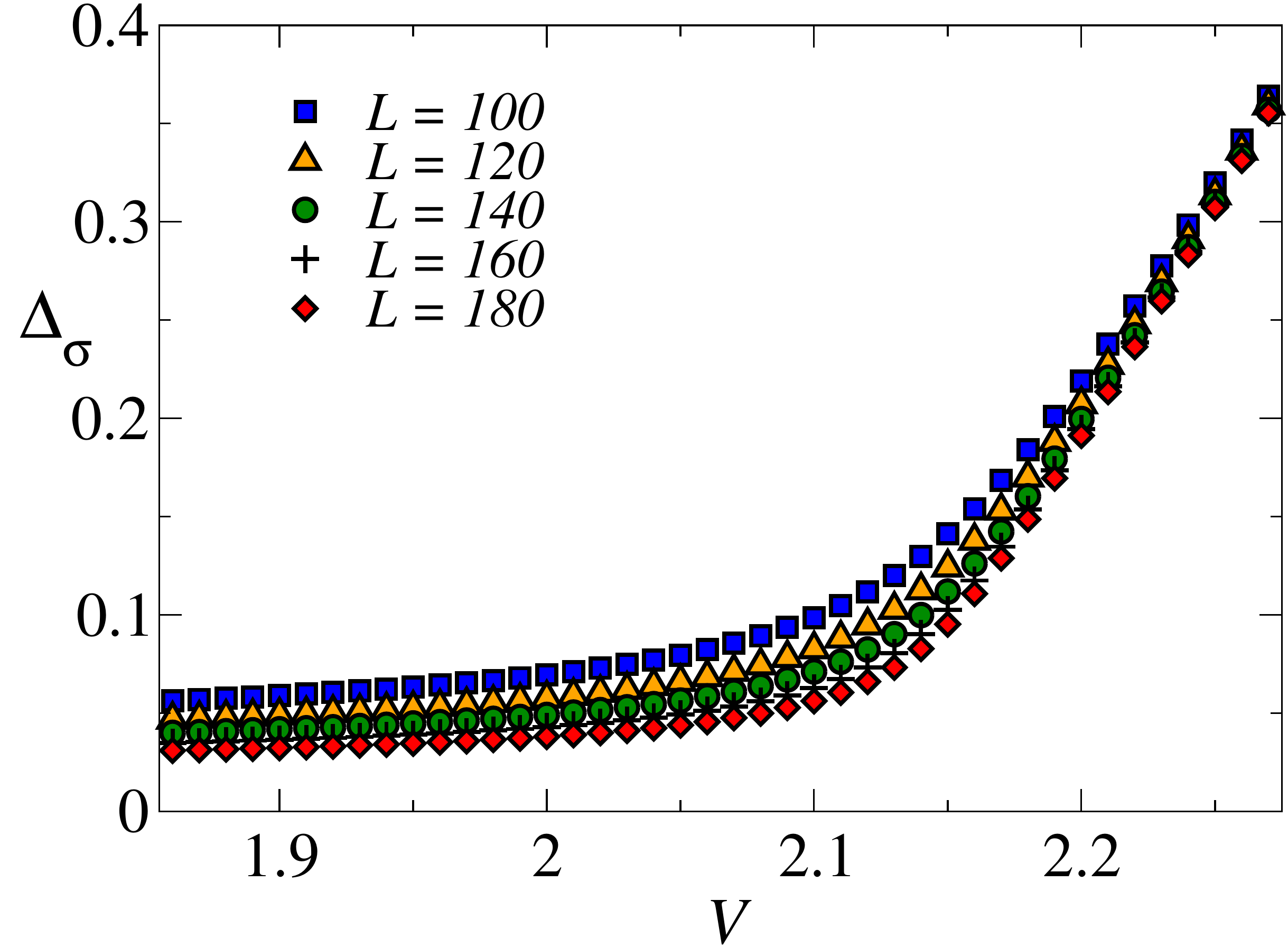}
 \caption{(Color online) Spin gaps for the EHM as functions of $V$ for different values of $L$, with $U=4$ and OBCs.}
 \label{EHM_Delta-V}
\end{figure}

We now extend our analysis of the gap-scaling method to multicomponent models, where the interplay between 
different energy scales can make the pinning down of BKT transitions even more complex than 
in single-species models. The first example we consider is the 1D extended Hubbard model (EHM), defined 
by the Hamiltonian:
\begin{equation}
\begin{split}
\hat{H}=&-t\sum_{j,\sigma}\left(\hat{c}_{j,\sigma}^\dagger \hat{c}^{}_{j+1,\sigma}+\mbox{H.c.}\right)
         +U\sum_j\hat{n}_{j,\uparrow}\hat{n}_{j,\downarrow}+\\
        &+V\sum_{j}{\hat{n}_{j}\hat{n}_{j+1}},
\end{split}\label{eq:hub}
\end{equation}
where $\hat{c}^{}_{j,\sigma}$ (with $\sigma=\uparrow$,$\downarrow$) is a spin-$1/2$ fermionic 
annihilation operator, $\hat{c}^{\dagger}_{j,\sigma}$ is its creation counterpart, 
and $\hat{n}_{j}=\sum_\sigma\hat{n}_{j,\sigma}$, with 
$\hat{n}_{j,\sigma}=\hat{c}^\dagger_{j,\sigma}\hat{c}^{}_{j,\sigma}$, is the site occupation operator;
$t$ is the hopping amplitude ($t=1$ sets our energy scale in what follows), $U$ is the on-site interaction 
coefficient, and $V$ parametrizes nearest-neighbor interactions. The phase diagram of this model has 
recently attracted quite some 
interest\cite{PhysRevLett.92.246404,PhysRevB.83.075111,PhysRevLett.99.216403,Tsuchiizu2004,Tam2006,Nakamura2000,
Jeckelmann2002,Barbiero:2013lq,Glocke:2007rz} due to the 
presence of a spontaneously dimerized phase supporting bond [or, more precisely, bond-charge-density-wave 
(BCDW)] order in the vicinity of the $U=2V$ line, with $U$, $V>0$. The BCDW phase 
intervenes between a charge-density-wave (CDW) and a spin-density-wave (SDW) phase, present, respectively, when 
$V\gg U$ and $U\gg V$. While, across this line, the low-energy charge sector of the theory remains gapped, 
the spin sector undergoes a BKT transition between SDW and BCDW at a critical value of $V$. 
This critical point has been debated, in particular, to discern whether a BCDW phase exists and, 
if it does, in which parameter regime.

Here, we apply the gap-scaling analysis to the spin gap at $U=4$ in order to detect the SDW to BCDW phase 
transition. The subsequent BCDW-CDW transition point is located around $V=2.16$.~\cite{PhysRevLett.99.216403}
The best estimates for the BKT between SDW and BCDW phases based on the finite-size scaling of the Luttinger 
parameter (under the assumption that logarithmic corrections are negligible) and on 
entanglement witnesses predict 
$V^\text{BKT}\simeq 1.88 - 2.02$.\cite{PhysRevLett.92.236401,PhysRevLett.99.216403,Mund:2009kq,Glocke:2007rz} We have 
performed DMRG simulations with chains up to $L=180$ sites (with OBCs), keeping up to 1024 states and up to 
8 finite-size sweeps in order to get truncation errors of order $10^{-7}$ (and a corresponding error in the 
spin gap of order $10^{-5}$). In Fig.~\ref{EHM_Delta-V}, we plot our results for the spin gap $\Delta_\sigma$
as a function of $V$, where $\Delta_\sigma(L) = E_{L, 0}(L) - E_{L,1}(L)$, with $E_{p, q}(L)$ is the ground state 
energy at size $L$ for a system with $p$ particles and magnetization $q$. For the smallest values of $V$ reported 
in the plot, $\Delta_\sigma$ decreases with increasing system size, while it does not seem to change with system size for 
the largest values of $V$ reported. This suggests that the spin gap closes at some $V_c$ in the thermodynamic 
limit, but does not quite help locating that point.

\begin{figure}[!t]
 \includegraphics[width=\columnwidth]{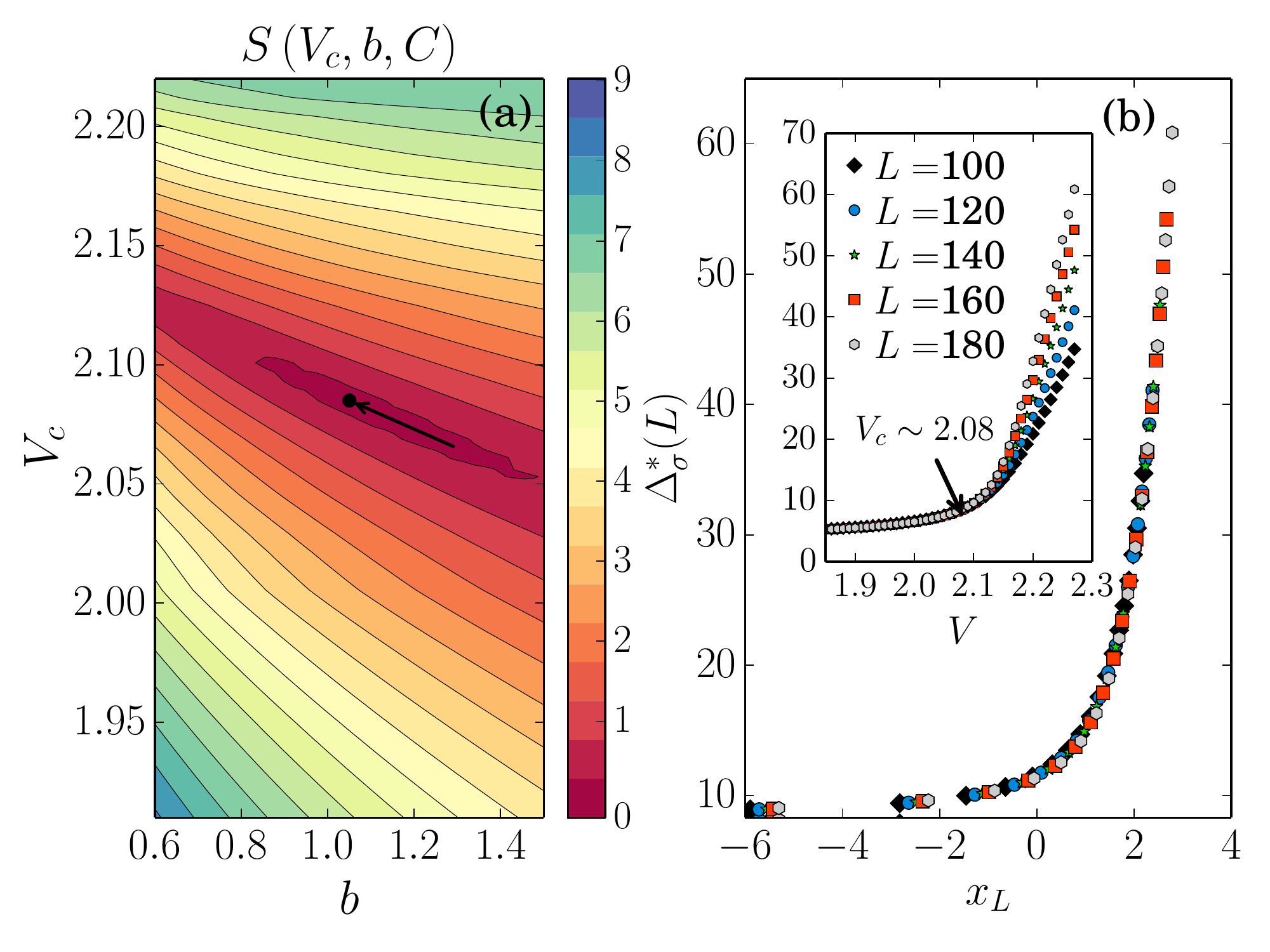}
 \caption{(Color online) (a) Contour plot of the sum of squared residuals $S(V_c,b,C)$ for the EHM model 
 with OBCs. (b) Best collapse of the data for $\Delta_{\sigma}^*(L)$ vs $x_L$ corresponding to $V_c=2.08$, 
 $b = 1.05$, and $C=-31$. The inset shows the rescaled gap vs $V$.}\label{EHM_sc}
\end{figure}

In Fig.~\ref{EHM_sc}, we summarize the results obtained through the gap-scaling procedure based on the data in 
Fig.~\ref{EHM_Delta-V}. The minimum of the function $S(V_c,b,C)$, displayed in Fig.~\ref{EHM_sc}(a), is located 
at $V_c=2.08\pm0.02$, $b=1.05\pm0.04$, and $C=-31\pm1$. The data produces the collapse presented in 
Fig.~\ref{EHM_sc}(b) and the inset shows that the data merge around $V\sim2.08$, in agreement with the critical 
value obtained from the minimization procedure.~\footnote{We have also checked that using the scaling ansatz 
proposed in Ref.~\onlinecite{Nakamura2000} gives results compatible with the present one at the 1\% level.} 
We note that our critical strength $V_c$ is above the estimates in 
Refs.~\onlinecite{PhysRevLett.99.216403,PhysRevLett.92.236401}, which means that a reduction of the 
size of the BCDW region is observed, in agreement with Ref.~\onlinecite{Glocke:2007rz}. Nevertheless, our 
estimate is still consistent with the presence of an 
intervening BCDW state in the phase diagram. We conclude that the gap scaling analysis, combined with numerical 
results on smaller chain sizes with respect to the ones employed in correlation-function and 
entanglement-witness studies, provides a rather accurate figure of merit for the phase transition 
point (at a one-percent level).

\section{One-dimensional anisotropic extended Hubbard model}\label{Sec:AEHM}

A simple variant of the EHM is its anisotropic version, the so called anisotropic extended Hubbard model 
(AEHM)\cite{Otsuka2000}
\begin{equation}
\begin{split}
\hat{H}=&-t\sum_{j,\sigma}\left(c_{j,\sigma}^\dagger c_{j+1,\sigma}+\mbox{H.c.}\right)
         +U\sum_j\hat{n}_{j,\uparrow}\hat{n}_{j,\downarrow}+\\
        &+V(1-\delta)\sum_{j,\sigma}{\hat{n}_{j,\sigma}\hat{n}_{j+1,\sigma}}+\\
        &+V(1+\delta)\sum_{j}\left(\hat{n}_{j,\uparrow}\hat{n}_{j+1,\downarrow}+
        \hat{n}_{j,\downarrow}\hat{n}_{j+1,\uparrow}\right),
\end{split}\label{eq:AEHM}
\end{equation}
whose main difference with respect to Eq.~\eqref{eq:hub} is in the last two terms, which reduce the original 
$SU(2)$ spin symmetry to $U(1)$, for any $\delta\neq 0$. We again set the hopping amplitude $t=1$ as our 
energy scale, and focus on the $\delta=0.2$ case. The phase diagram of this model has been explored by 
combining exact-diagonalization ($L\leq14$) and level-spectroscopy techniques~\cite{nomura1994}, and supports a finite 
bond-spin-density-wave (BSDW) for $U\lesssim 3$, intervening between a SDW and a CDW for $U\gg V$ and 
$V\gg U$, respectively, as in the EHM.\cite{Otsuka2000,PhysRevB.63.125111}

\begin{figure}[!b]
\includegraphics[width=0.75\columnwidth]{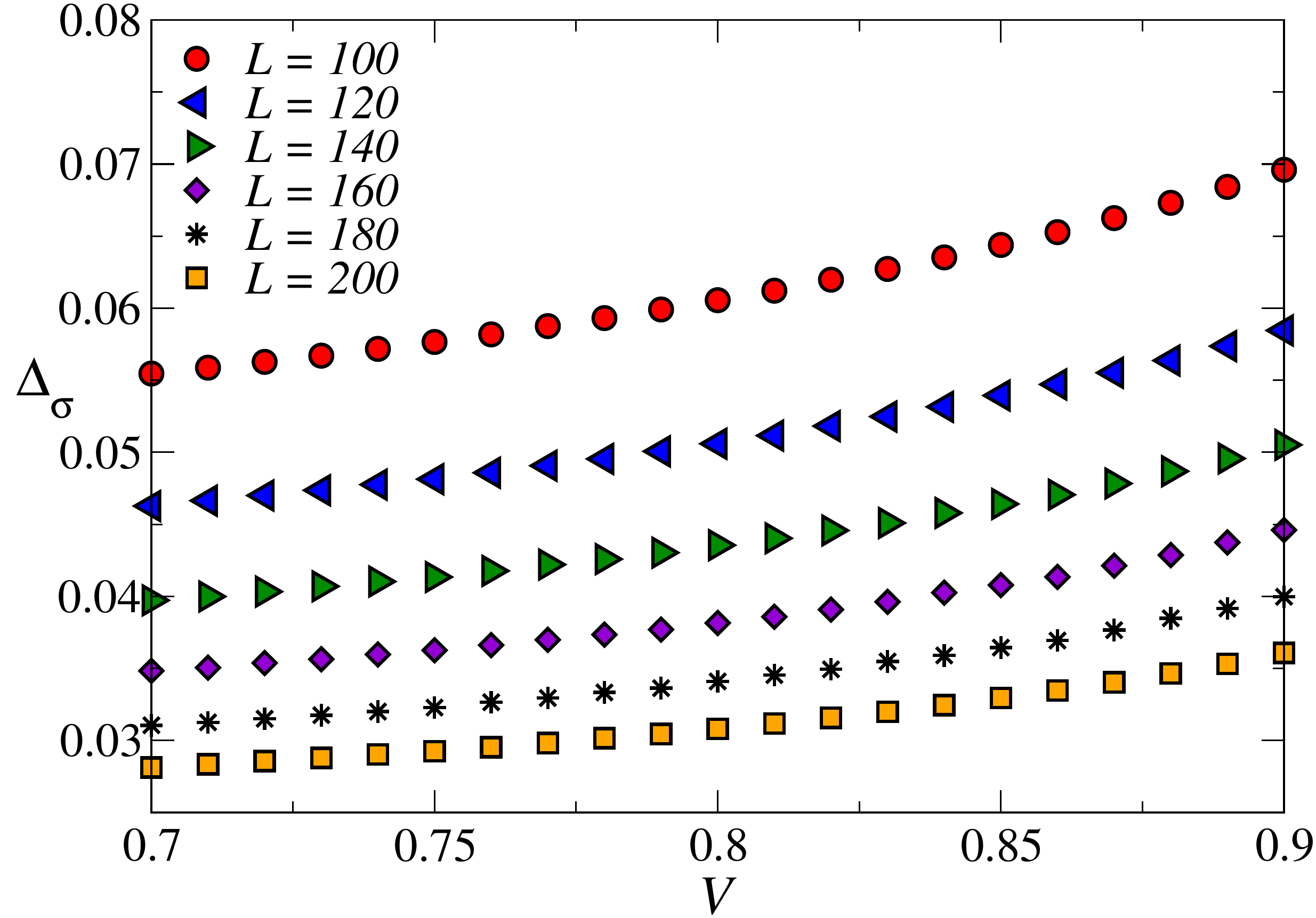}
\caption{(Color online) Spin gaps for the AEHM as functions of $V$ for 
different values of $L$, $\delta=0.2$, $U=1.5$, and OBCs.
}\label{AEHM_Delta-V_U1-5}
\end{figure}

Here, we are interested in the BKT transition separating the BSDW and the CDW for both 
$U=1.5$ and $U=2.5$. We determine the transition points by means of DMRG simulations of much larger system 
sizes than those accessible to exact diagonalization calculations.\cite{Otsuka2000} For the gap analysis, 
we have performed DMRG simulations in lattices with up to $L=200$ sites, keeping up to 512 states and up to 8 
finite-size sweeps in order to get truncation errors of order $5\times10^{-7}$ (and a corresponding 
error in the gap of order $10^{-5}$). The results are illustrated in Fig.~\ref{AEHM_Delta-V_U1-5}. 
In addition, we have evaluated the transition point by a complementary technique based on correlation 
functions that we describe below.

\begin{figure}[!t]
 \includegraphics[width=\columnwidth]{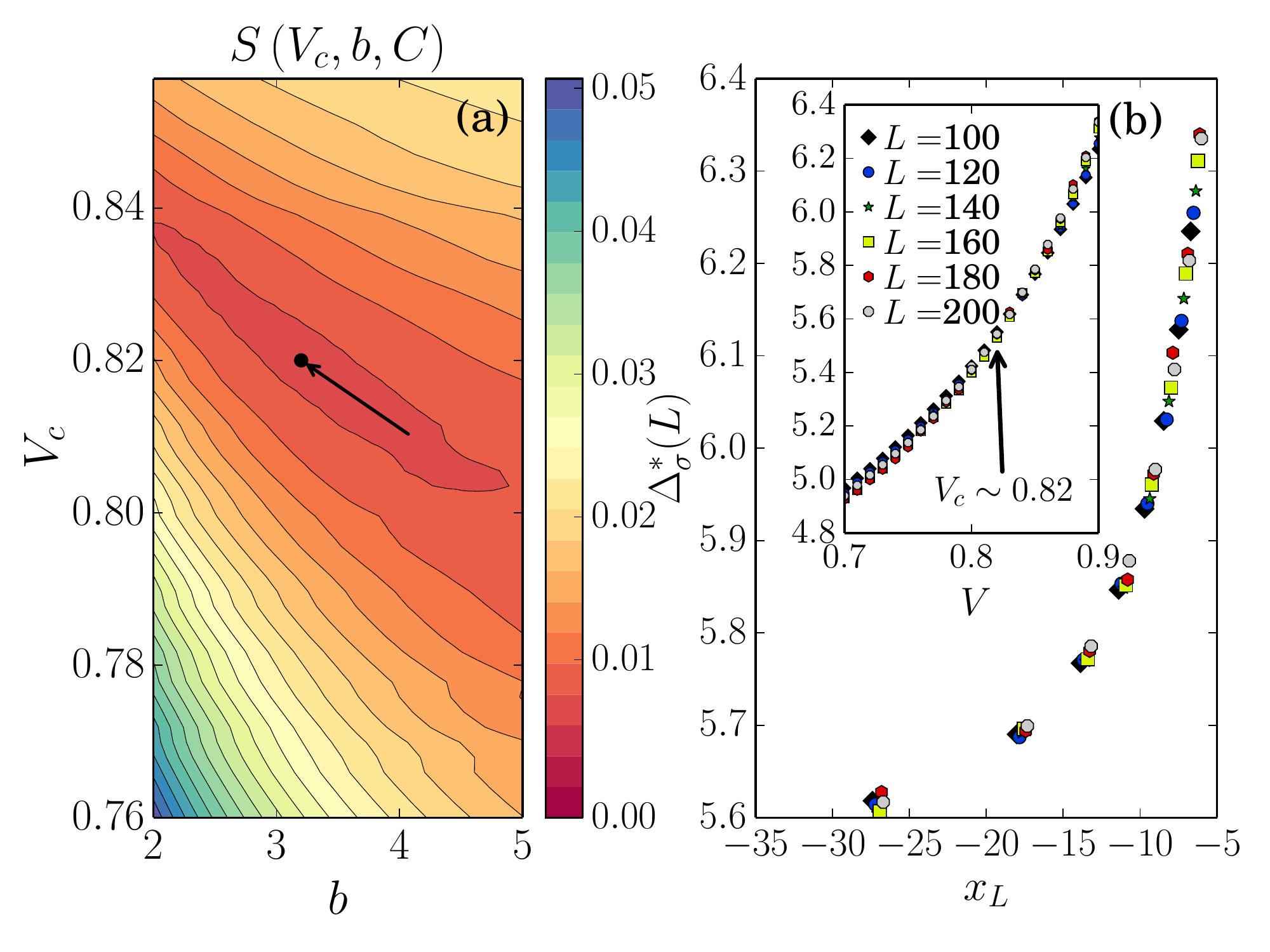}
 \caption{(Color online) (a) Contour plot of the sum of squared residuals $S(V_c,b,C)$ for the 
 AEHM model with OBCs, $U=1.5$ and $\delta=0.2$. (b) Best collapse of the data for $\Delta_{\sigma}^*(L)$ vs $x_L$ 
 corresponding to $V_c=0.82$, $b = 3.2$, and $C=-18.8$. The inset shows the rescaled gap vs $V$. }\label{AEHM_sc}
\end{figure}

In Fig.~\ref{AEHM_sc}, we report the results of our scaling analysis based on the data in 
Fig.~\ref{AEHM_Delta-V_U1-5}. The minimum of the function $S(V_c,b,C)$, displayed in 
Fig.~\ref{AEHM_sc}(a), is located at $V_c=0.82\pm0.03$, $b=3.2\pm0.1$, and $C=-18.8\pm0.2$. 
The data produces the collapse presented in Fig.~\ref{AEHM_sc}(b) and its inset shows a region 
around $V\sim0.82$ where the data merge, as expected from Eq.~\eqref{eq:scaling}. The same procedure 
applied to $\delta=0.2$ and $U=2.5$ yields $V_c=1.14\pm0.02$, $b=7.0\pm0.5$, and $C=-49\pm0.5$. 
These critical parameters are in agreement with the phase diagram from level spectroscopy 
measurements presented in Refs.~\onlinecite{Otsuka2000,PhysRevB.63.125111}.  
However, our estimates are extracted from much larger system size data, and, hence, we expect our results to be more accurate.

\subsection{Transition point from the spin Luttinger parameter}

\begin{figure}[!t]
 \includegraphics[width=0.75\columnwidth]{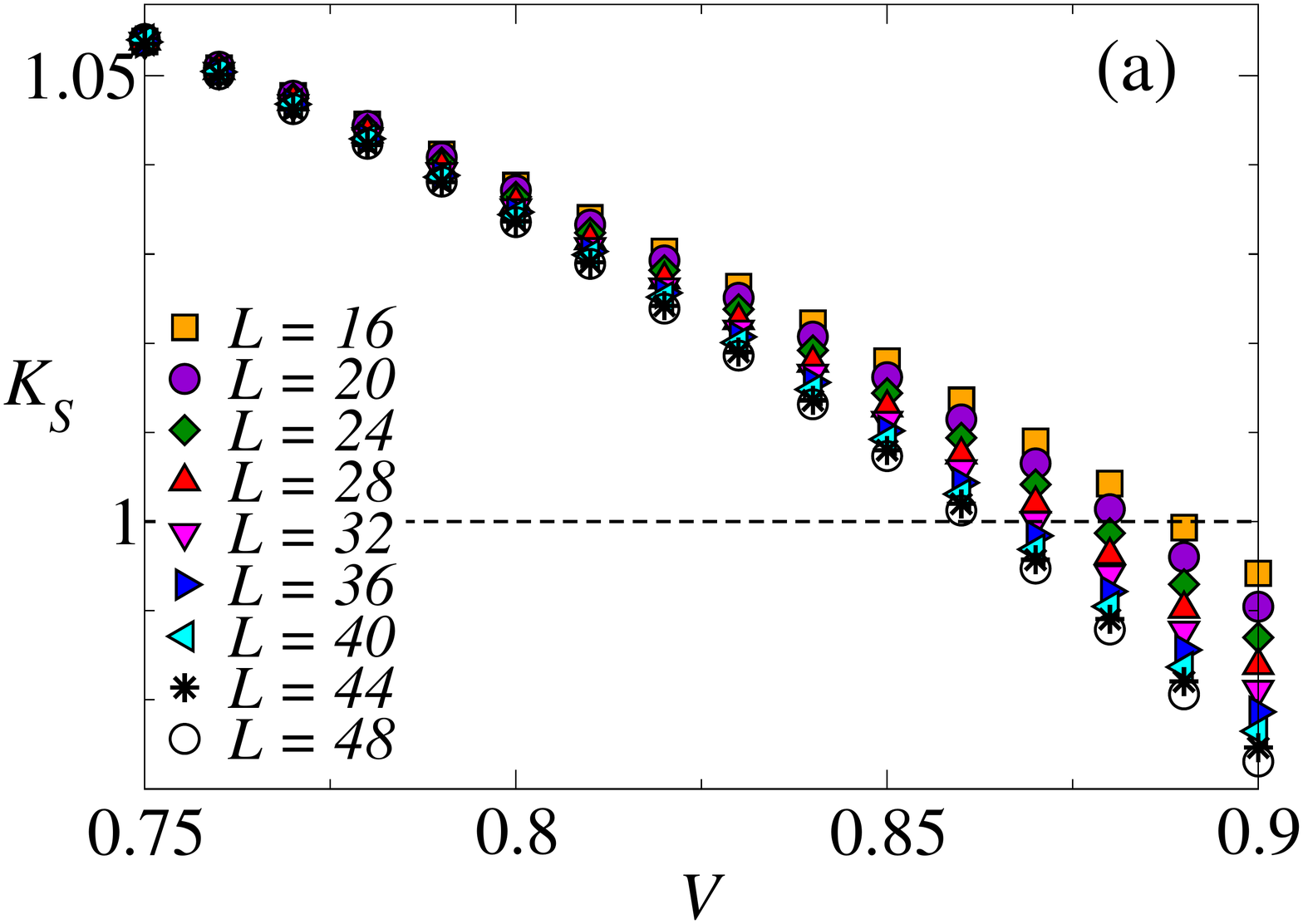}
 \includegraphics[width=0.75\columnwidth]{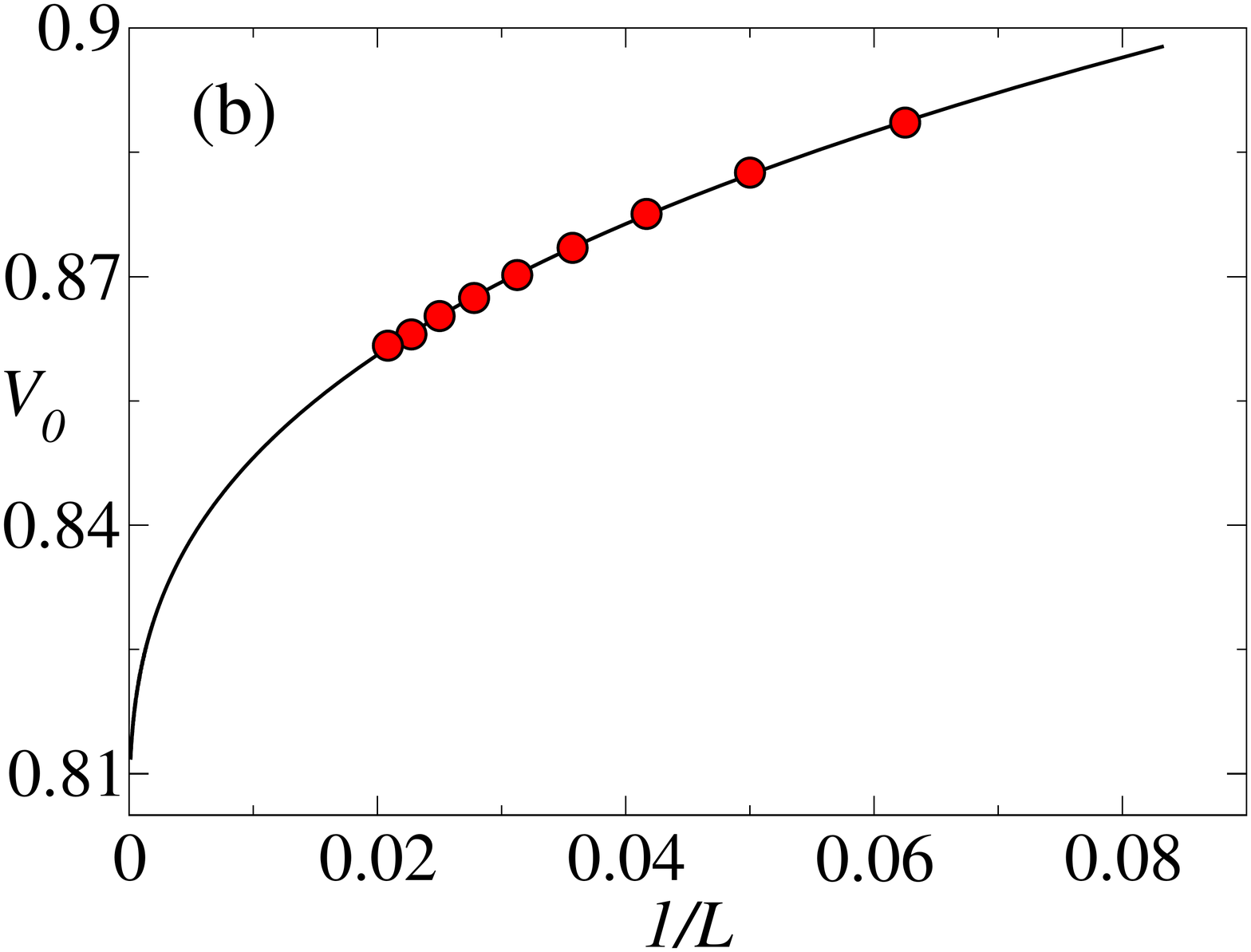}
 \caption{(Color online) (a) Spin Luttinger parameter as a function of $V$ for different system sizes for the AEHM. The 
 dashed line, $K_S=1$, depicts the value of $K_S$ at the critical point. (b) Scaling of the value of $V$ at which
 $K_S=1$ for each system size vs $1/L$. All results reported are for $U=1.5$ and $\delta=0.2$.}\label{Lpar}
\end{figure}

In order to have a quantitative benchmark for the gap-scaling analysis for this model, we have 
investigated the location of the BKT transition between BSDW and CDW using correlation-functions methods 
based on the underlying field-theoretical structure.\cite{Otsuka2000} At the transition point, the spin
Luttinger parameter flows to the BKT separatrix, that is, $K^*_S = 1$. In finite size samples, 
it is possible to extract $K_S(L)$ by monitoring the behavior of the spin structure factor:
\begin{equation}
S_s(k) = \frac{1}{L}\sum_{j, \ell} e^{ik(j-\ell)}(\langle S^z_jS^z_\ell \rangle -\langle S^z_j\rangle \langle S^z_\ell \rangle ),
\end{equation}
with $S^z_j = (\hat{n}_{j,\uparrow} - \hat{n}_{j,\downarrow} )/ 2$, and applying the relation:~\cite{Giamarchi2003}
\begin{equation}\label{Ks}
K_S(L) = L\frac{S_s(2\pi/L)}{2},
\end{equation}
which stems from the low-momentum behavior of the structure factor in a gapless phase, 
$S_s(q)\simeq qK_S/\pi$.\cite{PhysRevB.59.4665} For each system size, taking the smallest numerically available 
$q=2\pi/L$, this leads to Eq.~\eqref{Ks}. In order to avoid edge effects, we have performed DMRG 
simulations on samples with anti-periodic-boundary conditions~\footnote{For $L=4n, n\in\mathbb{N}$, 
we observe stronger finite-size effects for comparable system sizes, possibly due to the fact that the SDW phase 
becomes frustrated.} 
for various system sizes up to $L=48$, using up to $10$ finite-size sweeps and $1800$ states per 
block. The truncation and energy error were kept smaller than $10^{-5} 
(5\times10^{-5})$ for $L\leq 40$ ($L>40$). Single-site expectation values (such as 
$\langle\hat{n}_{j,\sigma}\rangle$) were found to be translationally invariant up to $10^{-7}$ 
corrections at most. Results for $K_S(V)$ vs $V$ for different system sizes are reported
in Fig.~\ref{Lpar}(a).

For each system size, we fit the function $K_S(V)$ with a fourth-order polynomial, and determine 
the value of $V_0(L)$ such that $K_S(L; V_0) = 1$ [point at which the curves for $K_S(V)$ vs $V$
cross the dashed line in Fig.~\ref{Lpar}(a)]. A finite-size-scaling analysis is then carried out 
on $V_{0}(L)$ in order to extract the critical value of $V$ in the thermodynamic limit 
by assuming the scaling form
\begin{equation}
V_0(L) = V_c + a_0 L^{-a_1},
\end{equation}
and performing a fourth-order polynomial fit using both least-square and Nelder-Mead methods. 
In addition, we have performed a linear fit using sizes $L>20$ for comparison. The results are 
illustrated in Fig.~\ref{Lpar}(b). For $U=1.5$, we find that $V_c = 0.81\pm 0.04$, 
where the error is estimated by comparing the fitting procedure using different sets of system sizes and different 
fitting techniques.

The critical point for $U=1.5$ obtained using the Luttinger-liquid parameter is consistent with the one 
from the gap-scaling analysis. However, the accuracy achieved in the latter is superior to that of 
the former approach. This because the Luttinger-liquid-based approach involves: 
(i) a fit and an extrapolation, and (ii) smaller system sizes, because of the need to compute correlations avoiding 
boundary effects, than those available in the gap scaling analysis. Furthermore, the gap scaling analysis accounts for logarithmic 
corrections [see Eq.~\eqref{eq:scaling}], which are difficult to incorporate in the scaling of the Luttinger 
parameter.  

\section{Conclusions and outlook}\label{Sec:concl}

We have studied various BKT transitions by means of a recently introduced gap-scaling analysis. 
Starting with the spin-3/2 XXZ model, where the critical anisotropy is known to be $J_z=1$, we 
ascertained the validity of the gap scaling procedure. Using both PBCs and OBCs, we found excellent 
agreement between the numerical results and the analytical one. We have shown that the scaling 
ansatz in Eq.~\eqref{eq:scaling} describes well the critical behavior of the gap data on finite 
systems, as observed from the quality of the data collapses presented in Figs.~\ref{32XXY_sc}, 
\ref{EHM_sc}, and \ref{AEHM_sc}. For the first time, we have successfully applied the 
gap scaling methodology to extended Hubbard models, where the interplay between different energy 
scales in the system makes the determination of critical parameters much more difficult than 
in spin models. In both the EHM and AEHM, our results are consistent with, but in principle more accurate 
(as they systematically include logarithmic corrections) than, 
previous estimates obtained using other techniques. We should stress that the critical parameters 
reported here for the transitions in the AEHM are the first to be obtained since early exact 
diagonalization results in small system sizes,\cite{Otsuka2000,PhysRevB.63.125111} which are usually 
affected by large finite-size effects. We have also compared our results 
from the gap scaling analysis with those from a method based on the determination of the Luttinger 
liquid parameter across the transition. They were found to be ingood agreement, but the gap 
scaling analysis is more accurate.

We stress that the gap scaling analysis discussed here offers significant advantages over 
other methods to detect BKT transitions used in the literature. First, the scaling ansatz 
in Eq.~\eqref{eq:scaling} includes logarithmic corrections to the gap, which are generally 
significant in BKT transitions.\cite{Eckle1997,PhysRevB.51.6163} Second, this methodology can 
be indistinctly applied to transitions involving the closing of either a charge or a spin gap, 
i.e., one can equally well study problems involving spins, fermions, bosons, multicomponent 
systems, etc.. This constitutes an advantage over well established techniques such as level 
spectroscopy, which, e.g., are hardly applicable to bosonic models~\cite{Carrasquilla2013} where the Hilbert space grows
extremely fast as a number of components, preventing an accurate finite-size scaling analysis.
Third, the gap is a quantity that, for large system sizes, can be obtained in 
different unbiased computational techniques, such as DMRG and quantum Monte Carlo approaches. 
Fourth, in DMRG (and usually in quantum Monte Carlo) simulations, the energies used in the 
determination of the gaps are variational, i.e., they are bounded and their quality can be 
easily assessed.

The demonstration of the aforementioned generality paves the way toward additional studies
of models whose location of a BKT transition is still debated. With comparable computational 
resources as the ones employed here, one could investigate the so-called asymmetric Hubbard 
model,\cite{Cazalilla:2005ly,Barbiero:2010qy} where a
BKT transition has been predicted separating a two-channel LL phase and a SDW in the repulsive regime, but 
where numerical and analytical approaches predicted different transition point 
locations.\cite{Cazalilla:2005ly,Roscilde2012} Moreover, a computationally more
demanding application could be the identification of different pairing regimes in three- and four-component 
Hubbard models. There, in the absence of SU(N) symmetries, a rich pairing pattern has been numerically and 
analytically put forward.\cite{Capponi:2008qy} However, a precise estimate of the transition lines is challenging 
due to strong spin-charge mixing, and as such, the gap scaling method could potentially serve as an unbiased 
estimate for the transition between the different pairing regimes in case
an exact BKT nature can be proven.

\section{Acknowledgments}
We thank C. Degli Esposti Boschi and F. Ortolani for help with the DMRG code. M.D. was supported by the ERC Synergy 
Grant UQUAM, SIQS, and SFB FoQuS (FWF Project No.~F4016-N23). J.C. acknowledges support from 
the John Templeton Foundation. Research at Perimeter Institute is supported through Industry Canada 
and by the Province of Ontario through the Ministry of Research \& Innovation. L.T. acknowledges financial 
support from the EU integrated project SIQS. E.E.  acknowledges the INFN grant QUANTUM  for partial financial 
support. M.R. was supported by the National Science Foundation, Grant No.~PHY13-18303.  

\bibliography{Bibliov2}
\end{document}